\documentstyle[12pt]{article}
\title{NONPERTURBATIVE EFFECTS IN QCD AT $T>0$}
\author{ Yu.A.Simonov\\
Institute of Theoretical and Experimental Physics\\ Moscow, 117259}
\date{}
\begin{document}
\maketitle

\begin{abstract}

The background field formalism is used to implement nonperturbative QCD
contributions into diagrammatic technic at $T>0$. The leading terms both in
the confining and nonconfining phase are identified at large $N_c$ and the
transition temperature is calculated in this limit, which appears to be in
good agreement with the Monte Carlo calculations.

\end{abstract}

\section{Introduction}

It is believed that the perturbative QCD is applicable in the deconfined
phase at large enough temperatures $T$, where the effective coupling
 constant $g(T)$ is small [1], while at small $T$ (in the confined phase)
 the nonperturbative effects instead are most important. However even at
 large T  the physics is not that simple: some effects, like screening
 (electric gluon mass), need a resummation of the perturbative series [2],
 while the effects connected with the magnetic gluon mass demonstrate the
 infrared divergence of the series [3].

 There is a substantial amount of data from lattice calculations in
 the deconfined phase, which can be explained  only by nonperturbative
 effects (see discussion below in section 6). Therefore one can  assume that
 nonperturbative effects are of vital importance both for small and large
 $T$ and should be taken into account before any dynamical perturbative
 scheme is applied. We suggest to use for this purpose
 the background field formalism [4], in connection
with the cluster expansion for background fields [5] and the Feynman-Schwinger
representation
(FSR) [6] for the Green's function.

This method has been used successfully for QCD at $T=0$ and  it contains
confinement
 as an essential ingredient, appearing through the area law of Wilson loops in
FSR.

It is a purpose of this paper to apply the method to the case of $T>0$ by
proper
modifications of FSR and the background field method (BFM). For this purpose
we write in Section 2 the familiar BFM expressions for the free energy with the
only
important new element: we do not consider the background field as classical
(as it is usually done) but rather make an average over an ensemble of
background fields, so that the latter enter final expressions for the free
energy  in the form of gauge-invariant correlators.

As a consequence the simple loop diagram-- the leading term $0(g^0)$ of the
usual perturbation theory --- becomes a set  of diagrams representing
multiple loops interacting via nonperturbative correlators. This is done in
Section 3 for gluons and in Section 4 for quarks.

The important new element there is a new form of FSR for a gluon or a quark
Green's
function appropriate for $T>0$: it contains a path integral winding on a
torus which compactifies the fourth Euclidean direction.  This enables us to
compute gauge-invariant Green's function for hadrons and response functions
for $T>0$.

Higher order amplitudes are discussed in Section 5. Here the behaviour of  the
QCD coupling constant $g$ in strong
background field at all distances  is exploited [7].

In Section  6 we identify the leading contributions to the free energy in the
limit when $N_c \rightarrow \infty$ and $g$  is small. It is gratifying that
the phase transition of the first order occurs already in this
oversimplifying case and the resulting transition temperature $T_c$ does not
depend on $N_c$ when $N_c$ is large.

Clearly the mechanism of the phase transition is the evaporation of a part of
the
gluonic condensate into a gas of almost free gluons [8].
The predicted value of $T_c$ agrees well with Monte Carlo calculations both in
the
case of gluodynamics [9] and in the case of QCD [10] for different number of
flavours.

In conclusion
possible prospectives of the proposed method are outlined.

\section{Basic  equations}

We start with standard formulas of the background field formalism [4]
generalized to the case of nonzero temperature. We assume that the gluonic
field $A_\mu$
 can be split into the background field $B_{\mu}$  and the quantum field
$a_{\mu}$
\begin{equation}
A_{\mu}= B_{\mu}+a_{\mu},
\end{equation}
both satisfying periodic boundary conditions
\begin{equation}
B_{\mu}(z_4, z_i) = B_{\mu}(z_4+n\beta, z_i), a_{\mu} (z_4, z_i) =
a_{\mu} (z_4+ n\beta, z_i),
\end{equation}

where $n$ is an integer and $\beta= 1/T$.

The partition function can be written as $$ Z(V,T,\mu=0) =<Z(B)>_B\;,$$
\begin{equation}
Z(B)=N\int D\phi exp (-\int^{\beta}_0 d\tau \int d^3x L(x,\tau))
\end{equation}
 where $\phi$ denotes all set of fields $a_{\mu}, \Psi, \Psi^+,N$ is a
normalization  constant, and the sign $<>_B$  means some averaging over
(nonperturbative) background fields $B_{\mu}$, the exact form of this averaging
is not needed for our purposes. Furthermore, we have
$$L(x,\tau)=\sum^{8}_{i=1} L_i,$$
where
\begin{eqnarray}
\nonumber
L_1=\frac{1}{4} (F^a_{\mu\nu}(B))^2;  L_2=\frac{1}{2} a_{\mu}^a W_{\mu\nu}^{ab}
a_{\nu}^b,
\\
L_3=\bar{\Theta}^a (D^2(B))_{ab}\Theta^b; L_4=-ig\bar{\Theta}^a (D_{\mu},
a_{\mu})_{ab}\Theta^b
\\
\nonumber
L_5=\frac{1}{2}\alpha (D_{\mu}(B)a_{\mu})^2;  L_6=L_{int} (a^3,a^4)
\\
\nonumber
L_7=- a_{\nu} D_{\mu}(B) F_{\mu\nu}(B); L_8=\Psi^+(m+\hat{D}(B+a))\Psi
\end{eqnarray}

Here $\bar{\Theta},\Theta$ are ghost fields, $\alpha$- gauge--fixing constant,
$L_6$
contains 3--gluon-- and 4--gluon vertices, and we keep the most general
background
field $B_{\mu}$, not satisfying classical equations, hence the
 the appearence of $L_7$.

 The inverse gluon propagator in the background gauge is
\begin{equation}
W^{ab}_{\mu\nu} =- D^2(B)_{ab} \cdot \delta_{\mu\nu} - 2 g F^c_{\mu\nu}(B)
f^{acb}
\end{equation}
where
\begin{equation}
(D_{\lambda})_{ca} = \partial_{\lambda} \delta_{ca} - ig T^b_{ca} B^b_{\lambda}
\equiv
\partial_{\lambda} \delta_{ca} - g f_{bca} B^b_{\lambda}
\end{equation}

We cosider first the case of pure  gluodinamics, $L_8\equiv 0$.

Integration over ghost and gluon degrees of freedom in (3) yields
\begin{eqnarray}
\nonumber
Z(B) =N'(det W(B))^{-1/2}_{reg} [det (-D_{\mu}(B)
D_{\mu}(B+a))]_{a=\frac{\delta}{\delta J}} \times
\\
\times \{ 1+ \sum^{\infty}_{l=1} \frac{S_{int}^l}{l!} (a= \frac{\delta}{\delta
J}) \}
exp (-\frac{1}{2} J W^{-1}J)_{J_{\mu}= \;\;\;\;\;D_{\mu}(B)F_{\mu\nu}(B)}
\end{eqnarray}

One can consider  strong background fields, so that $gB_{\mu}$ is large (as
compared to $\Lambda^2_{QCD}$), while $\alpha_s=\frac{g^2}{4\pi}$
in that strong background is small at all distances [7].

In this case Eq. (7) is a perturbative sum in powers of $g^n$,
arising from expansion in $(ga_{\mu})^n$.

In what follows we shall discuss the Feynman graphs for the free energy $F(T),$
connected to $Z(B)$ via
\begin{eqnarray}
F(T) = -T ln <Z(B)>_B
\end{eqnarray}

As will be seen, the lowest  order graphs already  contain  a nontrivial
dynamical mechanism for  the deconfinement transition, and those will be
considered in the next section.

\section{ The lowest  order gluon contribution}

To the lowest order in $ga_{\mu}$ ( and keeping all dependence on $gB_{\mu}$
explicitly) one has
\begin{eqnarray}
Z_0 = e^{-F_0(T)/T}= N'<exp(-F_0(B)/T)>_B
\end{eqnarray}
where using (7) $F_0(B)$ can be written as
\begin{eqnarray}
\nonumber
\frac{1}{T}F_0 (B) =\frac{1}{2} ln det W - ln det (-D^2(B))=
\\
=Sp\{ -\frac{1}{2} \int^{\infty}_{0}\zeta(t) \frac{dt}{t} e^{-tW}
+ \int^{\infty}_{0} \zeta (t)\frac{dt}{t}e^{tD^2(B)}\}
\end{eqnarray}

In (10) $Sp$ implies summing over all variables (Lorentz and color indices and
coordinates), $\zeta(t) = lim \frac{d}{ds} \frac{M^{2s}t^s}{\Gamma(s)}
\mid_{s=0}$
is a regularizing factor, one can use also the Pauli-Villars form for
$\zeta(t)$.

Graphically the first term on the r.h.s. of (10) is shown in Fig.1 and is a
gluon
loop in the background field, the second term is a ghost loop, shown in Fig.2.

Let us turn now to the averaging procedure in (9).  With the
notation $\varphi =-F_0(B)/T$, one can exploit in (9) the
cluster expansion [5] \begin{eqnarray} \nonumber <exp \varphi>_B = exp
\sum^{\infty}_{n=1} \ll \varphi^n \gg \frac{1}{n!}= \\ exp \{ <\varphi>_B
+\frac{1}{2} [<\varphi^2>_B - <\varphi>^2_B] + 0(\varphi^3) \}.
\end{eqnarray}
 To get a closer look at $<\varphi>_B$ we first should discuss termal
propagators of gluon and ghost in the background field.
We start with  the  thermal ghost propagator and write the FSR for it [6]
\begin{eqnarray}
(-D^2)^{-1}_{xy} =
  <x \mid \int^{\infty}_0 dt e^{tD^2(B)} \mid y> = \int^{\infty}_{0} dt
(Dz)^w_{xy} e^{-K}\hat{\Phi}(x,y)
\end{eqnarray}

Here $\hat{\Phi}$ is the parallel transporter in the adjoint representation
 along the trajectory of the ghost:
\begin{eqnarray}
\hat{\Phi}(x,y) = P exp (ig \int \hat{B}_{\mu}(z) dz_{\mu}),
\end{eqnarray}
also $$ K=\frac{1}{4} \int^t_0 d \tau \dot{z}^2_{\mu};\;\;\;\; \dot{z}_{\mu} =
\frac{\partial z_{\mu}(\tau)}{\partial \tau}$$
and $(Dz)^w_{xy}$ is a path integration with boundary conditions
imbedded (this is marked by the subscript $(xy)$) and with all possible
windings
in the Euclidean temporal direction.

One can write it explicitly as
\begin{eqnarray}
(Dz)^w_{xy} = \prod^{N}_{m=1} \frac{d^4 \zeta (m)}{(4\pi\varepsilon)^2}
\sum_{n=0,\pm,...} \frac{d^4p}{(2\pi)^4} e^{ip (\sum^{N}_{m=1} \zeta
(m) -(x-y) -n\beta \delta_{\mu 4})}
\end{eqnarray}
Here $\zeta (k) = z(k)-z(k-1), N\varepsilon =t$.

One can check that in the free case, $\hat{B}_{\mu}=0$, Eq.(12) reduces to
well-known
form of the free propagator
\begin{eqnarray}
\nonumber
(-\partial^2)^{-1}_{xy} = \int^{\infty}_0 dt e^{-\sum^{N}_1
\frac{\zeta^2(m)}{4\varepsilon}}
\prod_{m} \overline{ d\zeta (m)} \sum_{n} \frac{d^4p}{(2\pi)^4}
e^{ip(\Sigma \zeta (m)-(x-y)-n\beta\delta_{\mu 4})}=
\\
= \sum_{n} \int^{\infty}_0 e^{-p^2t-ip(x-z)-ip_4n\beta} dt
\frac{d^4p}{(2\pi)^4}
\end{eqnarray}

with $$ \overline {d\zeta (m)} \equiv \frac{d\zeta (m)}{(4 \pi
\varepsilon)^2}.$$

 Using the Poisson summation formula
\begin{equation} \frac{1}{2\pi} \sum_{n=0,\pm 1,\pm 2...} exp(ip_4 n\beta) =
\sum_{k=0,\pm 1, ...} \delta( p_4\beta -2\pi k ) \end{equation} one finally
gets the Jackiw-Templeton's form [1,2] \begin{equation}
(-\partial^2)^{-1}_{xy} = \sum_{k=0,\pm 1,...} \int \frac{Td^3p}{(2\pi)^3}
\frac{e^{-ip_i(x-y)_i- i2\pi k T ( x_4-y_4)}}{p^2_i+(2\pi kT)^2}
\end{equation}

Note that as expected the propagator (12),(17) corresponds to a sum  of ghost
paths with
all ppossible windings around the torus. The momentum integration in (14)
asserts that the sum of all infinitesimal "walks" $\zeta(m)$ should be equal to
the distance $ (x-y)$ modulo $n$ windings in the conpactified fourth
coordinate.

For the gluon propagator in the backgrounvd gauge we obtain similarly to
(12)
\begin{equation}
 (W)^{-1}_{xy}= \int^{\infty}_0 dt (Dz)^w_{xy} e^{-K} \hat{\Phi}_F(x,y)
\end{equation}
where
\begin{equation}
\hat{\Phi}_F(x,y)=P_F P exp(-2i g\int^t_0 \hat{F}(z(\tau))d\tau) exp ig
\int^x_y
\hat{B}_{\mu}dz_{\mu}
\end{equation}
 and the operators $P_F P$ are used to order insertions of $\hat{F}$ on the
trajectory
of the gluon [11,12].

Now we come back to the first term in (11), $<\varphi>_B$, which can be
represented with the help of (12) and (18) as
\begin{equation}
<\varphi>_B = \int \frac{dt}{t} \zeta (t) d^4x (Dz)^w_{xx} e^{-K}[\frac{1}{2}
tr
<\hat{\Phi}_F(x,x)>_B -<tr\hat{\Phi}(x,x)>_B]
\end{equation}
where the sign $tr$ implies summation over Lorentz and color indices.

In the Appendix 1 we show that Eq.(20) yields for $B_{\mu} =0$ the
usual result of the free gluon gas:
\begin{equation}
F_0 (B=0) =-T\varphi(B=0) = -(N^2_c -1)V_3 \frac{T^4\pi^2}{45}
\end{equation}

As a next example  we consider the case of nonzero component $\hat{B}_4$, while
$\hat{B}_i=0, i=1,2,3$. We also neglect for simplicity the difference between
$\hat{\Phi}_F$
and $\hat{\Phi}$, i.e. we put $\hat{F}=0$. Actually the interactions of
$\hat{F}$
 represent the interaction
of the gluon  spin with the background field [11, 12] and gives rise to
short-range
correlations (in the confining phase it also yields spin-orbit interaction, the
nonperturbative part of which, the Thomas precession, is long range, $\sim
1/r$)
with the notation
\begin{equation}
<\hat{\Omega}>_B = < exp ig \int^{\beta}_0 \hat{B}_4 (z) dz_4>_B
\end{equation}
one obtains from (20)
\begin{eqnarray}
\nonumber
 <\varphi>_B = Tr_c \int \frac{dt}{t} \zeta (t) V_3\beta
\frac{d^4p}{(2\pi)^4} \sum_{n=0,\pm 1} e^{-p^2t -i p_4
n\beta}<\hat{\Omega}^n>_B=
\\
=-Tr_cV_3 \int \frac{d^3p}{(2\pi)^3}\int^{M}_{\beta p} d\Theta (1+2\sum_{n=1,2}
e^{-\Theta n}<\hat{\Omega}^n>_B)=
\\
\nonumber
=Tr_c(-2V_3\int\frac{d^3p}{(2\pi)^3} <ln (1-e^{-\beta p} \hat{\Omega})>_B=
\\
=Tr_c<\frac{\beta V_3}{3\pi^2} \int^{\infty}_0 \frac{p^3 dp}{e^{\beta
p}\hat{\Omega}^{-1} - 1}>_B=
Tr_c \frac{2T^4}{\pi^2}V_3 \sum_{n=1,2,...}\frac{<\hat{\Omega}^n>_B
+<\hat{\Omega}^{*n}>_B}
{2n^4}
\end{eqnarray}

In the case $\Omega \equiv 1$ we come back to the free case, Eq.(21), since
\begin{equation}
\sum^{\infty}_{n=1} \frac{1}{n^4} = \frac{\pi^4}{90}
\end{equation}
The expression (24) for a constant field $\hat{B}_4$ yields the well-known
result
\begin{equation}
F_0(\hat{B}_4 = const) =- \frac{2V_3T^4}{\pi^2} Tr_c \sum_{n=1,2,...}
\frac{\cos(g \hat{B}_4 n\beta)}{n^4}
\end{equation}
where $Tr_c$ means trace over color indices.

One can notice in (26), (24) that in general the presence of field
$\hat{B}_4$(or $\hat{\Omega} \not= 1 $) descreases the modulus of $F_0$, i.e.
is not advantageous from the point of view of the minimum of $F_0$.

Thus $<\varphi >_B=-F_0/T$ in the deconfined regime represents a loop
contribution
of the gluon gas in the background field. Now we turn to the
confining regime and again calculate $<\varphi>_B$. It is  convenient to choose
another point $y$ on the loop in addition to the initial point $x$, as shown in
Fig.3
and to write $<\varphi>_B$ as
\begin{equation}
<\varphi>_B = \int^{\infty}_{0} \zeta (t) \frac{dt}{t} \int^t_0 \frac{dt_1}{t}
d^4y
d^4x (Dz)^w_{xy} (Dz')^w_{xy} e^{-K-K'-\sigma_a S}
\end{equation}
where we have introduced the area law for the closed contour average
\begin{equation}
<\hat{\Phi}(x,x)>= exp (-\sigma_a S)
\end{equation}
It is important at this point to stress that in the confining regime the total
contour
$C'(x,y,x)$  shown in Fig.4, made by trajectories $z(\tau)$ and $z'(\tau)$
in (27), should be a closed  contour, or in other words, in the representation
\begin{eqnarray}
\nonumber
Dz(\tau)=\prod^{N}_{m=1} \overline{d\zeta(m)} \sum_{n} \frac{d^4p}{(2\pi)^4}
exp(ip(\sum\zeta (m) -(x-y) -n\beta \delta_{\mu 4}))
\\
Dz'(\tau')=\prod^{N'}_{m'=1} \overline{d\zeta(m')} \sum_{n'}
\frac{d^4p'}{(2\pi)^4}
exp(ip'(\sum\zeta'(m') -(x-y) -n'\beta \delta_{\mu 4}))
\end{eqnarray}
one should choose
\begin{equation}
n=n'
\end{equation}
In this case  both trajectories $z(\tau)$ and $z'(\tau')$ are winding
in the same way on the torus and the minimal surface $S$ lies on the surface of
the
torus, winding around it $n$ times. In the case $n\not= n'$ there appear
a piece of length $ \mid n-n'\mid$ where a gluon is winding and  contributing
 an
infinite amount to the free energy
(since we assume that the free energy of an isolated
 quark or gluon is ininite).
Thus for $n=n'$ one can integrate in (27) , (29) over $d^4(x-y)$ as follows
(note
that $x_4-y_4$ changes in the limits $[0\beta]$)
\begin{equation}
d^4 (x-y) \sum_{n} exp [-i (\vec{p}+\vec{p'}) (\vec{x}-\vec{y}) - i (p_4+p'_4)
(x_4-y_4+n\beta)]= (2\pi)^4 \delta^4(p+p')
\end{equation}
We obtain for $<\varphi>_B$
\begin{equation}
<\varphi>_B= \int\zeta (t) \frac{dt}{t} \frac{dt_1}{t} V_3\beta G(t_1,t-t_1)
\end{equation}
where we have introduced
\begin{eqnarray}
G(t_1,t-t_1) = \frac{d^4p}{(2\pi)^4}e^{ip\sum\zeta_r} \prod \overline{d\zeta_r}
\overline{d\zeta_R} e^{-\frac{\sum\zeta^2_R(m)}{4\varepsilon} -
\frac{\sum\zeta^2_r(m)}
{4\varepsilon} -\sigma_a S}
\\
\mbox{and}\;\;\;
\nonumber
\zeta_r = \zeta(m) - \zeta'(m), \zeta_R=\frac{1}{2}(\zeta(m)+\zeta'(m))
\end{eqnarray}
One can notice that temperature enters only as a factor $\beta$ in (32) and it
is
cancelled in the free energy
\begin{equation}
F_0 = -T<\varphi>_B = - V_3 \int \zeta (t)\frac{dt}{t}\frac{dt_1}{t}
G(t_1,t-t_1)
\end{equation}
One can also rewrite $G$ through the two-gluon Green's function\\
 $G(xx,00,t,t-t_1)$, where gluons start at the point $0$ and meet at the
point $x$:  \begin{equation} G(t_1,t-t_1) \equiv \int d^4x
 G((xx;00;t_1,t-t_1).  \end{equation} In the free  case it is equal to
$\frac{d^4p}{(2\pi)^4} e^{-p^2t}$.  Note however that this result coincides
with the loop diagram for $T=0$, which is entirely absorbed by the
subtraction constant in calculating (21).

In case of confinement one obtains instead
\begin{equation}
G=\sum_{n}\varphi^2_n(0)e^{-M^2_n t}
\end{equation}
where $\varphi_n(0), M_n$ are wavefunction at origin and glueball mass
respectively.

In any case our result for the gluon loop with confinement, Eq.(34), does
not depend on temperature and should be included in the value of the free
energy
for zero temperature, which as we shall argue below is given by the scale
anomaly [14]
,i.e. by the gluonic condensate.

Note that physically this correspondence is quite understandable, since the
gluon
loop contributimon (34) is expressed through the two-gluon bound state at the
total
 momentum zero, i.e. through the condensate of bound gluon-gluon pairs
(glueballs), and
the latter is most generally given  by the gluonic condensate
$<F_{\mu\nu}F_{\mu\nu}>$ via the scale anomaly expression.

\section{The lowest order quark contribution}

Integrating over quark fields in (3) is done trivially and leads to the
following
additional factor in (7)
\begin{equation}
det (m+ \hat{D}(B+a)) = \frac {1}{2} det (m^2-\hat{D}^2(B+a))
\end{equation}
where we have used the symmetry property of eigenvalues of $\hat{D}$ In the
lowest approximation we omit $a_{\mu}$ in (37) and write the free
energy contribution $F_0(B)$ similary to the gluon
contribution (9-10) as
\begin{equation}
\frac{1}{T}F_0^q(B) =- \frac {1}{2} ln det (m^2-\hat{D}^2(B))=
-\frac{1}{2} Sp \int^{\infty}_{0} \zeta(t) \frac{dt}{t} e^{-tm^2+t\hat{D}^2(B)}
\end{equation}
where the sign $Sp$ has the same meaning as in (10) and
\begin{equation}
\hat{D}^2 =(D_{\mu} \gamma_{\mu})^2 = D^2_{\mu}(B) - gF_{\mu\nu}\sigma_{\mu\nu}
\equiv D^2-g\Sigma F; \;\;\; \sigma_{\mu\nu}
=+\frac{i}{4}(\gamma_{\mu}\gamma_{\nu}-
\gamma_{\nu}\gamma_{\mu});
\end{equation}
Our aim now is to exploit the FSR to represent (38)
in a form of the path integral, as it was done for gluons in (14). The
equivalent
form for quarks must implement the antisymmetric boundary conditions pertinent
to
fermions. It is easy to understand that the correct form for quark is
\begin{equation}
\frac{1}{T} F^q_0 (B) = -\frac{1}{2} tr \int^{\infty}_0 \zeta(t) \frac{dt}{t}
d^4x
\overline{ (Dz)}^w_{xx} e^{-K-tm^2}W_{\sum}(C_n)
\end{equation}
where $W_{\sum} (C_n) = P_F P_A exp ig \int_{C_n} A_{\mu} dz_{\mu}exp g(\sum
F)$
\begin{equation}
\overline {(Dz)}^w_{xy} = \prod^{N}_{m=1}\frac{d^4\zeta(m)}{(4\pi
\varepsilon)^2}
\sum_{n=0,\pm 1,\pm 2,...} (-1)^n\frac{d^4p}{(2\pi)^4} e^{ip(\sum^{N}_{m=1}
\zeta(m)-(x-y)-n\beta \delta_{\mu 4}}
\end{equation}

One can easily check that in the case $B_{\mu}=0$ one is recovering the
well-known
expression for the free quark gas
\begin{equation}
F^q_0(free\; quark)=-\frac{7\pi^2}{180} N_c V_3 T^4\cdot n_f,
\end{equation}
where $n_f$ is the number of flavors.
The derivation of (42) starting from the path-integral form(40) is done
similarly to the gluon case given in the Appendix 1.

The loop $C_n$ in (40) corresponds to $n$ windings in the  fourth direction.
Above the deconfinement transition temperature $T_c$ one  can visualize in (40)
the
appearance of the factor
\begin{equation}
\Omega = exp\; ig \int^{\beta}_0 B_4 (z) dz_4
\end{equation}
(Note the absence of the dash sign in (43) as compared to (22) implying that in
(43)
$B_4$ is taken in the fundamental representation). For the constant field $B_4$
and
$B_i=0, i=1,2,3$ one obtains
\begin{equation}
<F> =-\frac{V_3}{\pi^2} tr_c \sum^{\infty}_{n=1} \frac{\Omega^n+\Omega^{-n}}
{n^4} (-1)^{n+1}
\end{equation}

This result coincides with the obtained in the literature [13].

\section{Higher order contributions}

 We unify under this title two different types of contributions: i) higher
order
 cumulants from the cluster expansion (11) for the determinantal term (9); ii)
terms of
 the higher order in $g$, appearing in the perturbative expansion (7).

 We discuss first the terms i). A generic case is given by the quadratic
cumulant
 $\ll\varphi^2\gg_B $ which can be written as (we neglect the gluon spin terms)
\begin{equation}
\ll\varphi^2\gg_B=\int^{\infty}_0 \zeta (t)\frac{dt}{t}\int^{\infty}_0\zeta(t')
\frac{dt'}{t'} d^4x d^4x' e^{-K-K'} (Dz)^w_{xx}(Dz')^w_{x'x'}\times
\end{equation}
$$\times \ll tr \hat{\Phi}(x,x) tr \hat{\Phi}(x',x')\gg $$
where $\hat{\Phi}$ is defined in (13). The cumulant in the  integral (45)
corresponds to the
interaction between two gluon loops, each loop being a gauge-invariant object.
Hence the interaction of the loops, given by the cumulant, is suppressed as
$1/N^2_c$
both in the confining and nonconfining regimes(as compared to the contribution
$<\varphi>_B$ in
(11)).

   In the confining regime two gluons form a bound state -- a glueball.
   One can choose the initial glueball state connecting the loops $(xx) $ and
$(x'x')$
   by the stright line at some fixed $x_4=x'_4$, and then (45) can be rewritten
as
\begin{equation}
\ll\varphi^2\gg=\int \zeta (t)\frac{dt}{t}\int\zeta(t')
\frac{dt'}{t'} V^2_3 \beta^2  e^{-K-K'-\sigma_a S} (DR)^w_{xx}(Dr)^w_{00}
\end{equation}
   where $R$ and $r$ are central and relative coordinates of two gluons defined
in accordance
   with (33).
   The resulting integrals in (46) are similar to those in (32) with one
essential
   difference; in (32) the integration over all $d(x-y)$ as in (31) is
performed leading to the total
   momentum of two gluons $p+p'=0$.
   In contrast to that in (46) the initial and final central point $R$ is the
same and
   therefore all values of total momentum are possible as in the one--gluon
case, see (15)
   and Appendix 1.

   As a result one obtains the temperature dependence and the final result
   (see Appendix B for details)  is the free energy of the glueball state [15]
\begin{equation}
\ll\varphi^2\gg=\sum_{k} \frac{V_3(2m_kT)^{3/2}}{8\pi^{3/2}} e^{-m_k/T}
\end{equation}
where $m_k$ is the glueball mass in the glueball state $k$. Note that (47)  is
$0(N_c^0)$
since it is a color singlet contribution, in contrast to the gluon contribution
(21).

In the same way one can consider the next irreducible average,
$\ll\varphi^3\gg$.
The contribution is again a color singlet, corresponds to the three-gluon
glueball in the confining regime and is of the order $0(N_c^0)$.

The same features are common to higher order cumulants.

In a similar manner one can study the quark contributions to
$\ll\varphi^2\gg_B$.
In this case one has to replace in (45) the adjoint parallel transporters $
\hat{\Phi}$ by the fundamental ones and use the fermion boundary conditions in
$(Dz)^w_{xy}$ as in (41).
In the confining regime one obtains the meson contribution to the free energy
which has the
same form as in (47). It is of the order of $(N^0_c)$. The largest contribution
at small temperatures, $T\ll T_c$ comes from the pion gas, where neglecting the
pion
mass we have
\begin{equation}
F_0(pion \; gas) =- \frac{\pi^2}{30} V_3T^4
\end{equation}
One should note, that in the set i) introduced above the gluons interact whith
each
other via nonperturbative field $B_{\mu}$, which can be exemplified by the
correlators $\ll F_{\mu\nu}(B) F_{\lambda\sigma}(B)\gg$ etc. In contrast to
that in
the set ii) all perturbative exchanges are taken into account. These are
proportional
to  $g^n$ and in the usual treatment  $g=g(T)$ is
small at large temperature while it may diverge as $T\rightarrow \Lambda$ [1]
\begin{equation}
g^2(T) =  \frac{24\pi^2}{(11 N_c-2N_f) ln T/\Lambda},
\end{equation}

In the strong background field however one has an additional scale parameter
$<tr F^2_{\mu\nu}
(B)>\equiv B$ which defines dynamics both at small and large distances and
temperatures.
One can show that when $B\gg \Lambda^4$ this scale defines the value of the
coupling constant
$g(B,T)$ and $g(B,T)$ remains small even when $T$ tends to zero [7]. Therefore
one may hope that
the perturbative terms (set ii)) which yield relatively small contribution
($\sim
10 \%)$ at  $T\sim 3T_c$, remain of the same order at all temperatures.

\section{Calculation of the deconfiniment temperature in the leading order of
the $1/N_c$
expansion}

In this section we estimate the deconfinement temperature keeping
only leading terms in both phases -- confining and
deconfining. Results given below have been published in a short form in [8]. We
start with the zero temperature and note that the $NP$ energy density
$\varepsilon$ is connected via the scale anomaly [14] to the gluonic condensate
[16] (we
neglect the quark masses since only light quarks
 are important at low temperatures)
\begin{equation}
\varepsilon = \frac{\beta(\alpha_s)}{16\alpha_s}<G^a_{\mu\nu} G^a_{\mu\nu}>
\cong - \frac{11}{3} N_c \frac{\alpha_s}{32\pi}<G^2>
\end{equation}

One can associate with $\varepsilon$ the zero--temperature limit of the free
(Helmholtz)
energy $F(T=0)$ and write $F$ at nonzero temperature as
\begin{equation}
F=\varepsilon V_3+f(T)
\end{equation}
where $f(T)$ is to be computed using the formulas (8) and subsequent ones.
Since $f(T)$ is regularized by subtracting the zero-temperature limit $f(T=0)$
and all divergencies are present in the latter term, Eq.(51) is a definition of
the regularized limit $F(T=0)$ where only $NP$ contribution in the gluonic
condensate
is present. A check of correctness of our normalization of $F(T=0)$ is
given by the computation of the leading contribution in Section 3.

There the perturbative contribution (21) goes to zero as $T\rightarrow 0$ in
accordance with our prescription that perturbative contributions are normalized
to
vanish at $T\rightarrow 0$. If however one takes into account the confining
background,
one gets for $T\rightarrow 0$ a nonzero contribution (34), which is exactly due
to
a pair of gluons connected by the adjoint string (28) and with zero total
momentum (see
(31)).

This is exactly the contribution which one expects from a gluon pair forming a
condensate.

The low--temperature phase consists of the gluonic condensate (50), glueballs
(47), mesons
(we keep only the pion (48)) and their interacting conglomerates:
\begin{equation}
F_{low}=\varepsilon V_3-T\sum_{K}\frac{V_3(2m_KT)^{3/2}}{8\pi^{3/2}}e^{m_K/T}-
\frac{\pi^2}{30} V_3T^4+0(1/N_c)
\end{equation}
Here the first term is the leading in $N_c$; it grows as $N_c^2$ [17], while
the ideal
gas of glueballs and mesons contributes $0(N_c^0)$. Note that the interaction
between white  objects is suppressed at $N_c\rightarrow \infty $ [18] and is
presented in (52) by the last term.

The high--temperature phase is belived to be the gas of quarks and gluons,
interacting
perturbatively [1,2].
We shall now argue that i) there is at $T>T_c$ another component of the phase,
namely
a part of the gluonic condensate, which has not evaporated during the phase
transition; ii) there is a $NP$ interaction between quarks and gluons which is
important
in some situations.

To start we remind that the $NP$ interaction is governed by the correlators [6]
\begin{equation}
G_{\mu\nu,\lambda \sigma} \equiv<tr [F_{\mu\nu}(x)\Phi(x,y)F_{\lambda \sigma}
(y) \Phi(y,x)]>
\end{equation}
which at $T=0$ contain two independent Lorentz
structures with scalar coefficients $D(x-y)$ and $D_1(x-y)$ [5,6]
\begin{equation}
G_{\rho \mu,\sigma\nu}(u) =(\delta_{\rho \sigma}\delta_{\mu\nu} -
\delta_{\rho\nu} \delta_{\mu\sigma}) D(u) + \frac{1}{2}
[\frac{\partial}{\partial u_{\rho}} ((u)_{\sigma}\delta_{\mu\nu}-
(u)_{\nu} \delta_{\mu\sigma})+(\rho\sigma \leftrightarrow \mu\nu)] D_1(u)
\end{equation}
At $T>0$ the Lorentz invariance is broken and one obtains two independent
correlators-- for electric and magnetic fields respectively [19]
\begin{equation}
G_{ik}^{(E)} \equiv G_{i4, k4} =\delta_{ik} D^{(E)}(u) +\frac{1}{2} [...]
D^{(E)}_1 (u)
\end{equation}
\begin{equation}
G_{ik}^{(B)} \equiv G_{em, pq}\frac{1}{2}e_{iem}\frac{1}{2} e_{kpq} =
\delta_{ik} D^{B}(u) +\frac{1}{2} [...] D^{B}_1 (u) \end{equation}

For the gluon condensate one must put $u=0$ in (54-55) with the result
\begin{equation}
<tr F_{\mu\nu}(0)F_{\mu\nu}(0)> = D^E(0) + D_1^E (0) +D^B(0)+D_1^B(0)
\end{equation}
Out of four functions in (56) only one, $D^E$, is connected to the confinement,
since
the string tension $\sigma$ is expressed through it [5,6]
\begin{equation}
\sigma = \frac{1}{2} \int\int^{\infty}_{-\infty} D^E(\sqrt{u_1^2+u^2_4}) du_1
du_4 +...
\end{equation}
where dots imply contributions of higher-order cumulants.

Note that the integral (58) is confined to the region $
\sqrt{u^2_1+u_4^2}\preceq
T_g$, since $D^E$ falls off exponentially outside.
The value of $ T_g$ -- the gluonic
correlation length -- is the order of $0,2 fm$ [20].

Therefore for the temperatures $T\preceq T_c\simeq 200 MeV$ the periodicity in
the
fourth direction does not change $\sigma$ significantly, and
the deconfinement occurs  when $D^E$ disappears at some $T=T_c$
together with $\sigma$. There is no reason why $ D_1^E, D^B,D^B_1$ disappear at
$T=T_c$
or at $T>T_C$. Moreover, there is evidence [21] that $D^B$, which
defines the area law of the spacial Wilson loops, should be nonzero at $T>T_c$.
Additional evidence comes from the hadronic screening lengths [22],
also implying the confining dynamics in the spacial directions (we refer the
reader
to [9,19] for more discussion of this point).

Thus we can assume that only $D^E$ disappears at $T=T_c$. How gluonic
condensate changes
with temperature? The answer comes from lattice calculations [23] and shows
that
$<trF^2_{\mu\nu}(0)>$ is rather stable and only a part of it disappears
near $T=T_c$. This is understandable from the theoretical point of view, since
the
natural scale of $<tr F^2(0)>$ is given by the lowest glueball mass
approximately equal to the  mass of the dilaton [24] which is of the order of
$1 Gev$, and therefore $<tr F^2(0)>$
should not strongly change in the range $0\preceq T\preceq T_c$ . Thus we come
to the following
model for high--temperature phase:
\begin{equation}
F_{high} =(1-\eta)\varepsilon V_3-(N^2_c-1)V_3 \frac{T^4\pi^2}{45}-
\frac{7\pi^2}{180}N_cV_3T^4 n_f
\end{equation}
where
\begin{equation}
\eta = \frac{D^E(0)}{<trF^2(0)>}
,
\end{equation}

Since $D_1$ enters the $NP$ tensor force of quarkonia [11] one can make some
estimate
of $D_1/D:\;\;\;\; D_1/D \preceq 1;$ the lattice calculations [20] at $T=0$
yield $\frac{D_1(u)}{D(u)}\preceq\frac{1}{3}, u> 0.1 fm$.

Therefore it seems reasonable to neglect $D_1^E$ and $D_1^B$ ( note that
$D_1^E = D_1^B=D_1$ at $T=0$) and having in mind that $D^E=D^B$ at $T=0$,
one has an estimate \begin{equation} \eta \cong 1/2 \end{equation} As
another option with $D^E(0)=D^E_1(0)=D^B(0) =D^B_1(0)$, one can choose
$\eta=1/4$.

We have neglected in (59) i) higher order terms in $g^n$, ii) higher order
cumulants
$\ll\varphi^n\gg$ with $n\succeq 2$ and iii) the influence of the term $\Omega$
(24) and
(43). As we discussed above, we  believe that i) $\alpha_s$ is not large,
$\alpha_s \preceq
0.5$ at all temperatures ii) higher cumulants are suppessed  as $0(1/N_c^2)$,
iii) from the structure of (24) and (43) one can notice that  $\Omega \not= 1$
tends to decrease $\mid F\mid$ and since in the equilibrium the phase
chooses the minimal $F$ (maximal $\mid F \mid$) it will choose by that
$\Omega =1$.   Note also, that studies of $F$ for constant
$B_4$ [25] seem to support our conclusion (in lowest orders graphs).

Of course our argument here is very qualitative and can be used only for a
rough estimate of $T_c$. A more detailed study of the  role of $\Omega$ is
now in progress [26]. We now can plot both curves $F_{high}$ (59), $
F_{low}$ (52) or rather the pressure $P=-F/V_3$ as a function of temperature
and use again the principle of minimal $F$ (maximal $P$). One can see in
Fig.4 that there exists indeed some critical temperature $T=T_c$ and the
system prefers to be in the low phase at $T<T_c$ and in the high phase at
$T>T_c$. The transition is of the first order, and $T_c$ can be computed
equalizing (59) and (52)

In the leading order $0(N^2_c)$ one obtains:
\begin{equation}
T_c = (\eta \cdot \frac{45}{\pi^2}\cdot
\frac{11N_c}{N^2_c-1}\frac{\alpha_s}{96\pi}
<G^2>)^{1/4}
\end{equation}

We note that $T_c$ is $0(N_c^0)$ which is reasonable since strong interactions
(except
for baryons) have finite limit both in masses and the range of interaction when
$N_c\rightarrow \infty$. With the standard value [16].
\begin{equation}
G_2 \equiv \frac{\alpha_s}{\pi}<G^2> =0.012 GeV^4
\end{equation}
we have  for $\eta= 1/2$
\begin{equation}
T_c = (0.196 \eta G_2)^{1/4}=0.185 Gev
\end{equation}
However number of flavours $n_f=0$ gluonic gluonic condensate should be
2-3 times larger [16], which yields $T_c=0.22-0.24 GeV$ and  agrees well with
the lattice calculations for
gluodynamics [9].

In the next order $0(N_c)$ we take into account the quark contribution and
still disregard
all meson and glueball contributions (as well as nonideal gas corrections for
gluons
and quarks). We obtain
\begin{equation}
T_c =  \eta^{1/4}(\frac{\frac{11}{3} N_c \frac{\alpha_s<G^2>}{32\pi}}
{\frac{\pi^2}{45}(N^2_c-1)+ \frac{7\pi^2}{180}N_cn_f})^{1/4}
\end{equation}

For the choice (61) and $n_f=2$ we have $T_c=0.15 GeV$ while for $n_f=4$ we
have $T_c=0.134 Gev$, which agrees well with recent lattice calculations
[9,10] $T_c^{QCD}$(lattice)=$0.14 \sigma GeV$ and $0.131 GeV$ respectively.

\section{Conclusion}

We have started in the paper the background field
formalism for $T>0$ with the purpose to take into account
the NP contributions in a systematic way.

 The perturbation theory was formulated where expansion is of two different
 kinds: the cluster expansion for nonperturbative  interaction and the usual
 expansion in powers of the coupling constant for perturbative quarks and
 gluons in the strong NP background.

 In this first paper we concentrated on the lowest order terms: i) the
 gluonic condensate in the scale anomaly term ii) gluonic and quark loop
 diagrams in the background field.

 We have shown in the previous Section  that  already a rough
 approximtion, where only terms  i) and ii) are  kept, yields
 a reasonable picture of the deconfining phase transition  where
 a) the order of the transition b) the numerical value of $T_c$
 c) existence of non-perturbative effects at $T>T_c$, are correctly
 predicted. We also demonstrated that the terms i) and ii) are leading
 in the large $N_c$ limit and the resulting value of $T_c$
 doesn't depend  on $N_c$ in this limit.

 We plan to expand the formalism to include higher order terms and check
 the stability of our predictions. First of all, the role of gluonic
 (adjoint) Polyakov line $\Omega$ is being clarified [26]. This role  is
 especially important for the question of the latent heat, which comes out
 too large with $\Omega=1$ as compared to lattice calculations.

 Another important possibility of the presented
 formalism is the study of infrared singularities of the free energy for
$T>T_c$
 connected to difficulty with the magnetic mass [3,13].

 In the vaccum background field of QCD the confinement at $T>T_c$ is still
 present in the special planes [19] which cures the  infrared divergencies
 and can solve in principle the problem of magnetic mass.  This will be
 discussed in detail in a subsequent paper.
 Note added: after the original version of this manuscript has been
 published as a preprint [27], the author became aware of a similar estimate
 of $T_c$ in [28], where it was assumed that no nonperturbative
 configurations are present for $T>T_c$ in the high temperature phase. This
 corresponds to $\eta =1$ and results in $\sim 25$\% higher values of $T_c$
 as  compared to ours.

 However in the pictures of [28] it is difficult to explain lattice results
 of [21-22] which require nonperturbative magnetic configuration for
 $T>T_c$.

 The author is grateful for discussions to the members of the theoretical
 seminar of ITEP. A part of this work has been done while the author was a
guest of
 the Max-Planck Institut f\"{u}r Kernphysik in Heidelberg.
 It is a pleasure for him to thank Professor H.A.Weidenmueller and all the
staff of the
 Insitute for a kind hospitality. Fruitful discussions with H.G.Dosch,
 H.Leutwyler, P.Minkowski and H.D.Pirner are  gratefully acknowleged.

\newpage

\underline{\bf APPENDIX }\\

{\bf Calculation of the free energy of a
noninteracting gluon loop via\\
 the path integral, Eq.(20).}\\

\setcounter{equation}{0}
\def\theequation{A1.\arabic{equation}}

The square brackets in (20) yield $(\frac{1}{2}\cdot 4 -1) (N_c^2-1) =
(N^2_c-1)$.
The integral $(Dz)^w_{xx}$ can be calculated as the integral in $d\zeta(m) $
(see
Eq.(15)) and using also (16) we get
\begin{equation}
\frac{\varphi(B=0)}{N_c^2-1} = V_3 \beta \int^{\infty}_0 \frac{dt}{t} \zeta(t)
e^{-tp^2} dp_4 \frac{d^3p}{(2\pi)^3} \sum_{k} \delta (p_4\beta -2\pi k)=
\end{equation}
\begin{equation}
=\sum_{k} \int \frac{d}{ds} (\frac{M^2}{\vec{p}^2 +(2\pi k T)^2})^s \mid_{s=0}
\frac{V_3d^3p}{(2\pi)^3} = \sum_{k} \frac{V_3 d^3 p}{(2\pi)^3} ln
\frac{M^2}{\bar{p}^2+(2\pi k T)^2} =
\end{equation}
\begin{equation}
=\sum_{k=0,\pm 1,...} \frac{V_3d^3 p}{(2\pi)^3} [-\int^{\beta^2p^2}_1
\frac{d\Theta^2}
{\Theta^2+(2\pi k)^2} - ln(1+(2\pi k)^2) + ln M^2 \beta^2]
\end{equation}

Using a relation [1]
\begin{equation}
\sum_{n} \frac{1}{n^2+(\frac{z}{2\pi})^2} = \frac{2\pi^2}{z}(1+
\frac{2}{e^z-1})
\end{equation}

we obtain
$$\frac{\varphi}{N_c^2-1}=-\int
\frac{V_3d^3p}{(2\pi)^3} \int^{\beta\mid\vec{p}\mid}_1 d\Theta (1+
\frac{2}{e^{\Theta}-1})+ inf.
constants=$$
\begin{equation}
-\int
\frac{V_3d^3p}{(2\pi)^3}[\beta p +2ln (1-e^{-\beta p})+
const]=
\end{equation}
$$=\frac{\beta}{3\pi^2} \int^{\infty}_0 \frac{p^3dp V_3}{e^{\beta p}-1}=
\frac{\pi^2 T^3}{45};\;\;\;\; \mbox{or}\;\;\; F_0=-(N^2_c-1)
\frac{V_3T^4\pi^2}{45} $$
\newpage
\hspace{3cm}\underline{ FIGURE CAPTIONS}

\vspace{2cm}

  Fig.1. The gluon loop diagram in the  background field $B_{\mu}$ (the
  first term on the r.h.s. of Eq.(10)).

\vspace{1cm}

  Fig. 2. The ghost loop diagram in the background field $B_{\mu}$ (the
  second term on the r.h.s. of Eq. (10)).

\vspace{1cm}

  Fig. 3. The gluon loop with the confining film as a result of averaging
  over background field - Eq.(27).

\vspace{1cm}

  Fig. 4. Paths of integration $z(\tau)$ and $z'(\tau)$ in Eq.(27) in the
  in the confining regime.

\vspace{1cm}

  Fig.5. The pressure $P_{low}$ (corresponds to $F_{low}$ Eq.(52)) as a
  function of temperature - dash- dotted line. The pressure $P_{high}$
  corresponding to $F_{high}$, eq.(59)) with the quark contribution -solid
  line, without it - dashed line. Transition temperatures $T_c(q)$ and $T_c$
  obtained with and without quark contribution respectively.

\end{document}